\documentclass[aps,pra,superscriptaddress,floatfix,showpacs,10pt,twocolumn]{revtex4}
\usepackage[dvips]{graphicx}
\usepackage{amssymb}
\usepackage{amsmath}
\begin{document}
\title{Generation of pure, ionic entangled states via linear optics}
\author{Ming Yang}
\email{mingyang@ahu.edu.cn}
\author{Zhuo-Liang Cao}
\email{zlcao@ahu.edu.cn(Corresponding~Author)} \affiliation{ School
of Physics \& Material Science, Anhui University, Hefei, 230039,
People's Republic of China}
\begin{abstract}
In this paper, we propose a novel scheme to generate two-ion
maximally entangled states from either pure product states or mixed
states using linear optics. Our new scheme is mainly based on the
ionic interference. Because the proposed scheme can generate pure
maximally entangled states from mixed states, we denote it as
purification-like generation scheme. The scheme does not need a Bell
state analyzer as the existing entanglement generation schemes do,
it also avoids the difficulty of synchronizing the arrival time of
the two scattered photons faced by the existing schemes, thus the
proposed new entanglement generation scheme can be implemented more
easily in practice.
\end{abstract}
\pacs{03.67.Mn, 03.67.Hk, 03.67.Pp, 42.50.Dv}
\maketitle

\section{INTRODUCTION}
Quantum superposition principle is a fundamental principle in
quantum physics. When used in composite system, quantum
superposition principle can induce an entirely new result, which is
different from the classical physics, i.e. quantum
entanglement~\cite{en}. After a long time debate on the completeness
of quantum mechanics between Niels Bohr and Albert Einstein, it has
been generally accepted that the entanglement between two systems
exists. Entanglement state is the state that can not be expressed as
the product state of the two systems~\cite{en}. In this sense, the
quantum entanglement is used to disprove the local hidden variable
theory~\cite{hidden}. Because of the non-locality feature of
entanglement, entangled states have been widely used in quantum
information processing, such as quantum cryptography~\cite{cryp},
quantum computer~\cite{comput}, and quantum
teleportation~\cite{tele}. All of the above applications are based
on the entangled states, so the generation of entangled states plays
a critical role in quantum information processing. Many theoretical
and experimental schemes for the generation of entangled states have
been proposed in Cavity QED~\cite{generation}, ion trap~\cite{ion},
and NMR~\cite{nmr}. In photonic case, the polarization entangled
photons have been generated in experiment by using Spontaneous
Parametric-Down conversion~\cite{spdc}. For atomic case, the schemes
for the generation of entangled atomic states have been
proposed~\cite{generation}. Shi-Biao Zheng and Guang-Can Guo have
presented a realizable scheme for the generation of entangled atomic
states, which is mainly based on the dispersive interaction between
atoms and cavity modes. The obvious advantage of it is that the
cavity is only virtually excited during the process and the
requirement on the cavity quality is greatly loosened, which opens a
promising perspective for quantum information
processing~\cite{zheng}. This scheme has been realized in experiment
by the S. Haroche group~\cite{haroche}.

Alternatively, the entangled atomic states also can be generated via
atomic interference~\cite{ bose, browne, cabrillo, dlm, dlm1, feng,
lloyd, plenio, protsenko, simon}. Most of the schemes work as
follows: the two scattered(or leakage) photons from two spatially
separated atoms(or cavities) will be mixed by a Bell State
Analyzer(BSA) or Polarization Beam Splitter(PBS), and the two
photons will be detected after the BSA or PBS. Because we can not
distinguish from which atoms the two photons are scattered, the two
atoms will be left in entangled states after the photon detection.
The schemes of this type can entangle the spatially separated atoms,
but there is still a serious difficulty. These schemes require the
two photons reach the BSA or PBS simultaneously. Motivated by
Xing-Xiang Zhou's proposal on non-distortion quantum
interrogation~\cite{nqi}, we have proposed an entanglement
purification scheme for arbitrary unknown ionic states via linear
optics~\cite{mepra}. In this paper, we will propose a novel
entanglement generation scheme, which is free of the problem of
simultaneity. In our scheme,the photon wave function of one incident
circular polarized photon will be split into two parts, transmitted
part and the reflected part, by an ordinary nonpolarizing $50$--$50$
Beam Splitter(BS). Two multi-level ions will be pre-placed on the
two possible pathes of the photon. After interacting with the ions,
the two parts of the photon wave function will be re-combined by the
second ordinary nonpolarizing $50$--$50$ BS. Through detecting the
photon after the second BS, we can decide whether the entangled
ionic pairs has been created or not. The main part of the setup can
be regarded as a Mach-Zehnder interferometer(MZI). The BS is an
ordinary one, and the relative phase problem inherent in the
previous schemes has been avoided in our scheme. The photon detected
in this scheme is a circular polarized one, and can be detected
easier than the scattered one used in the previous schemes. It is
not easy to make the two photons from two different ions interfere
in the previous schemes. But in our scheme, the photon wave function
has been split into two parts in MZI, and the coherent condition is
satisfied naturally.

If nothing has been done on the entanglement generation setup before
the creation process, the probability of success is relative low
because of the low scattering rate of photon. So after discussion on
the original setup, we will make some modification on the main setup
to enhance the efficiency, i.e. the MZI will be surrounded by an
optical cavity resonant with the ionic transition. This cavity will
enhance the scattering rate and the successful probability of
entanglement generation. If we lengthen the two arms of the MZI, the
setup can entangle the two ions spatially separated, because the
polarization of photon can be preserved in a polarization-preserving
fiber over a long distance.

In the long distance case, two separate ions can be in a product
state or in a mixed state evolved from the pure entangled state
before distribution. If the two ions initially in mixed state are
placed on the setup, an pure maximally entangled state can be
extracted, and the efficiency of it can exceed the product state
case provided the mixed state satisfies some condition. In this
sense, the scheme for the mixed state case looks like a entanglement
purification process, but it is not the case, because entanglement
purification involves only classical communication and local
operations~\cite{purification}. So we only denote it as
"purification-like" generation scheme.

\section{GENERATION SCHEME FOR THE PRODUCT INITIAL STATES CASE}

\begin{figure}
\includegraphics[scale=0.4]{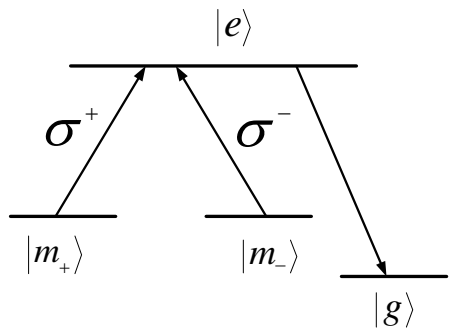}
\caption{\label{level}The level configuration of the ions. The ions
in $|m_{+}\rangle$ (or $|m_{-}\rangle$ ) can be excited into the
excited state $|e\rangle$ by absorbing one $\sigma^{+}\textrm{ (or
}\sigma^{-})$ polarized photon, and then it will decay to the stable
ground state $|g\rangle$ and scatter a photon rapidly. Here, we
assume that the decay process is so rapid that the probability of
excited emission can be neglected.}
\end{figure}

Next, we will discuss the entanglement generation process in
details. Here, we will consider two identical ions, and they are all
multi-level systems. The level configuration of the ions has been
depicted in Fig. \ref{level}.

\begin{figure}
\includegraphics[scale=0.7]{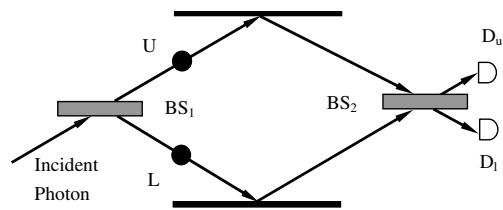}
\caption{\label{setup}The setup for the generation of two-ion
entangled states. $BS_{1}$ and $BS_{2}$ denote the two identical
nonpolarizing $50$--$50$ BSs. $U$ and $L$ denote the two ions on the
upper and the lower arm of the interferometer. $D_{u}$ and $D_{l}$
are two polarization sensitive single photon detectors at the output
upper and lower port.}
\end{figure}

Where $|m_{+}\rangle$ and $|m_{-}\rangle$ are two degenerate
metastable states which are used to store quantum information.
$|e\rangle$ is a excited state of ions and $|g\rangle$ is the stable
ground state. Ions in states $|m_{+}\rangle$ (or $|m_{-}\rangle$)
can be excited into the $|e\rangle$ state by absorbing one
$\sigma^{+}(\textrm{ or }\sigma^{-})$ circular polarized photon with
unit efficiency, then it will decay to ground state $|g\rangle$
rapidly and scatter a photon. This process can be expressed as:
\begin{equation}\label{scatter}
\hat{a}_{\pm}^{+}|0\rangle|m_{\pm}\rangle\longrightarrow|S\rangle|g\rangle.
\end{equation}
where $|S\rangle$ denotes the scattered photons which we assume will
not be reabsorbed by the ions and can be filtered away from the
detectors. Although this process does not always occur despite the
photon impinging on the ions, we still consider the ideal case to
demonstrate the process, and then we will consider how to enhance
the scattering rate by adding an optical cavity. The setup for
generation of maximally entangled ionic states is depicted in Fig.
\ref{setup}.

One MZI with two BSs is the main part of the generation setup. One
$\sigma^{+}$ polarized photon enters the MZI from the left lower
port. The MZI is initially adjusted (without ions) such that the
upper detector($D_{u}$) registers photons with certainty. In the
case of the existence of two ions in arbitrary superposition states
of $|m_{+}\rangle$ and $|m_{-}\rangle$ at each arm of the MZI(the
two ions can be placed on the two arms of the MZI by using the
trapping techniques~\cite{trapping}), the upper and the lower
detectors($D_{u} \textrm{and} D_{l}$) all have the probability of
fire. If we select the superposition coefficients of the initial
states of the two ions appropriately, we can get the maximally
entangled ionic states conditioned on the fire at $D_{l}$. Suppose
that the two ions ($U, L$) are initially prepared in the following
states:
\begin{subequations}
\begin{equation}
|\Psi\rangle_{U}=\alpha|m_{+}\rangle_{U}+\beta|m_{-}\rangle_{U},
\end{equation}
\begin{equation}
|\Psi\rangle_{L}=a|m_{+}\rangle_{L}+b|m_{-}\rangle_{L},
\end{equation}
\end{subequations}
where the coefficients $\alpha, \beta, a, b$ satisfy
$|\alpha|^{2}+|\beta|^{2}=1$ and $|a|^{2}+|b|^{2}=1$. These states
can be prepared by a laser pulse focused on the ion. The effect of
the BS on the input photon can be expressed as:
\begin{subequations}
\begin{equation}
\hat{a}_{\rightleftarrows,l}^{+}|0\rangle\stackrel{\textrm{BS}}{\longrightarrow}
\frac{1}{\sqrt{2}}(\hat{a}_{\rightleftarrows,u}^{+}\pm{i}\hat{a}_{\rightleftarrows,l}^{+})|0\rangle,
\end{equation}
\begin{equation}
\hat{a}_{\rightleftarrows,u}^{+}|0\rangle\stackrel{\textrm{BS}}{\longrightarrow}
\frac{1}{\sqrt{2}}(\hat{a}_{\rightleftarrows,l}^{+}\pm{i}\hat{a}_{\rightleftarrows,u}^{+})|0\rangle.
\end{equation}
\end{subequations}
That is to say, the BS takes no effect on the polarization of the
input photon, and reflects the wave function with a
$\pm\frac{\pi}{2}$ phase shift corresponding to the propagation
direction of the photon~\cite{nqi}.

Next, we will trace the input photon and give the evolution of the
total system. After one $\sigma^{+}$ polarized photon entering the
left lower port of the MZI, its wave function will be split into two
parts (the upper arm and the lower arm) by $BS_{1}$. Because the two
ions are placed on the two arms, they will interact with the
different parts of the wave function. Then the two parts of the wave
function will be combined by $BS_{2}$. The total evolution of the
system can be expressed as follow:
\begin{align}\label{evolution}
&\hat{a}_{\rightarrow,l,+}^{+}|0\rangle(\alpha|m_{+}\rangle_{U}+\beta|m_{-}\rangle_{U})
(a|m_{+}\rangle_{L}+b|m_{-}\rangle_{L})\nonumber\\
&\longrightarrow\frac{1}{\sqrt{2}}\alpha|S\rangle_{U}|g\rangle_{U}(a|m_{+}\rangle_{L}+b|m_{-}\rangle_{L})\nonumber\\
&+\frac{i}{\sqrt{2}}a|S\rangle_{L}|g\rangle_{L}(\alpha|m_{+}\rangle_{U}+\beta|m_{-}\rangle_{U})\nonumber\\
&+\frac{i}{2}\hat{a}_{\rightarrow,u,+}^{+}|0\rangle(\beta{a}|m_{-}\rangle_{U}|m_{+}\rangle_{L}
+\alpha{b}|m_{+}\rangle_{U}|m_{-}\rangle_{L}\nonumber\\
&+2\beta{b}|m_{-}\rangle_{U}|m_{-}\rangle_{L})\nonumber\\
&+\frac{1}{2}\hat{a}_{\rightarrow,l,+}^{+}|0\rangle(\beta{a}|m_{-}\rangle_{U}|m_{+}\rangle_{L}
-\alpha{b}|m_{+}\rangle_{U}|m_{-}\rangle_{L}).
\end{align}
From the above result, we can get that the two ions will be left in
three possible states corresponding to three measurement results on
the two output ports respectively. If the $D_{l}$ fires, we get
two-ion entangled states:
$\beta{a}|m_{-}\rangle_{U}|m_{+}\rangle_{L}
-\alpha{b}|m_{+}\rangle_{U}|m_{-}\rangle_{L}$. If we modulate the
coefficients of the initial states to make $\alpha, a, \beta, b$
satisfy $|\alpha|=|a|$ and $|\beta|=|b|$, the two ions can be left
in maximally entangled state
$|\Psi\rangle_{UL}=\frac{1}{\sqrt{2}}(|m_{-}\rangle_{U}|m_{+}\rangle_{L}
-|m_{+}\rangle_{U}|m_{-}\rangle_{L})$ with probability
$P=\frac{1}{2}|a|^{2}(1-|a|^{2})$ . From this analysis, we conclude
that the two ions must be prepared in the same superposition state
initially, then we can get the two-ion maximally entangled state.
The successful probability is a function of the modulus of the
initial states.

Here we only discussed the ideal case where the ion decays are
coherent. In fact, the ion decays will be essentially incoherent
and very few of the photons will be in the correct direction in
the free space, so a resonant cavity must be introduced for each
ion to achieve directional emission of the photons from each
ion~\cite{simon}. To simplify the figures, we do not draw these
cavities in the generation setups.

\begin{figure}
\includegraphics[scale=0.7]{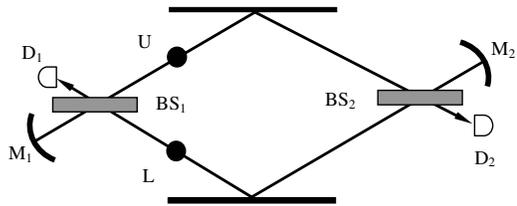}
\caption{\label{newsetup}The modified setup for entanglement
generation. $D_{1}$ and $D_{2}$ are two detectors. $M_{1}$ and
$M_{2}$ are two mirrors, which can form an optical cavity, i.e.
\emph{enclosure cavity}.}
\end{figure}

In addition, we have only discussed the ideal case where we
suppose a photon impinging on an ion always leads to the process
described by Eq.(\ref{scatter}). But in most cases the photon will
not be scattered by the ions. If the ions are placed inside the
MZI, it would mean that detector $D_u$ will most likely fire as
before but without any entanglement between the ions being
created. To enhance the scattering rate, an optical cavity will be
added to the MZI. This cavity encloses the MZI, and it is
different from the resonant cavities surrounding the ions, so we
denote the cavity enclosing the MZI as \emph{enclosure cavity}.
The modified setup is depicted in Fig. \ref{newsetup}.

Each side of the \emph{enclosure cavity} has a very high
reflection rate. One side($M_{1}$) is placed at the left lower
port of the MZI, the other($M_{2}$) at the right upper port. Thus,
if the photon has not been scattered by the ions, it will leave
MZI at the right upper port. It will be reflected into MZI by
$M_{2}$, and analogously for the description of $M_{1}$. That is
to say, if the photon is not scattered by the ions, it will
vibrate in the \emph{enclosure cavity} through MZI, which will
enhance the scattering rate naturally. From Eq. (\ref{evolution}),
if the photon has been scattered by one ion, there is still
possibility that one photon exit the MZI at the right upper port.
Then the same process will repeat until the $D_{1}$ or $D_{2}$
register one photon, which indicates the entanglement generation
process succeeds. During the process, if $D_{1}$ or $D_{2}$ do not
register photons after an interval of the order of the lifetime of
a photon, we must re-input one $\sigma^{+}$ polarized photon into
the MZI through the cavity \emph{enclosure cavity} side to repeat
the entanglement generation process until the $D_{1}$ or $D_{2}$
register one photon. To clarify the evolution process, we suppose
that one photon exiting the MZI at the right upper port indicates
the process of the third term of the Eq.(\ref{evolution})(this
assumption is optimal because the vibration of photon in the
cavity \emph{enclosure cavity} will enhance the chance of
occurring the scattering process of Eq. (\ref{scatter}). Once the
scattering process occurs the repetition of the third term of Eq.
(\ref{evolution}) will occur subsequently). After iteration of the
above process, the probability for ions in scattered states is
$\frac{1}{2}(|\alpha|^{2}+|a|^{2})+\frac{1}{6}(|\alpha{b}|^{2}+|\beta{a}|^{2})$,
the probability for ions in product state
$|m_{-}\rangle_{U}|m_{-}\rangle_{L}$ is $|\beta{b}|^{2}$, and the
probability for ions in entangled state
$\beta{a}|m_{-}\rangle_{U}|m_{+}\rangle_{L}
-\alpha{b}|m_{+}\rangle_{U}|m_{-}\rangle_{L}$ is
$\frac{1}{3}(|\alpha{b}|^{2}+|\beta{a}|^{2})$. If the initial
states of the two ions satisfy the condition $|\alpha|=|a|$ and
$|\beta|=|b|$, the two ions will be left in maximally entangled
state$|\Psi\rangle_{UL}=\frac{1}{\sqrt{2}}(|m_{-}\rangle_{U}|m_{+}\rangle_{L}
-|m_{+}\rangle_{U}|m_{-}\rangle_{L})$ with probability
$\frac{2}{3}|a|^{2}(1-|a|^{2})$. From the above result of
iteration, we get that the added optical cavity \emph{enclosure
cavity} enhance the scattering rate and the efficiency of
entanglement creation.

\section{GENERATION SCHEME FOR THE MIXED INITIAL STATES CASE}

Most of the previous preparation schemes prepare the entangled
states at one location, and then the entangled particles will be
distributed among different users for quantum communication purpose.
But during the transmission of the particles, it will unavoidably
couple with environments, and then the entanglement will degrade
exponentially. So the entangled states after distribution are
usually mixed ones, which need the purification process~\cite{mepra,
purification, pjw, distillation, distillation1} before use. We
consider the generation from two ions that are initially in mixed
state. For clarity, we will first give the evolution induced by the
original setup in Fig. \ref{setup}, then give some discussions on
the process if we use the modified setup in Fig. \ref{newsetup}.

Suppose that the initial mixed state is in the following
form~\cite{pjw,pjw1}:
\begin{equation}
\rho_{UL}=F|\Psi^{+}\rangle_{UL}\langle\Psi^{+}|+(1-F)|\Phi^{+}\rangle_{UL}\langle\Phi^{+}|.
\end{equation}
where $|\Psi^{+}\rangle_{UL}=\frac{1}{\sqrt{2}}
(|m_{+}\rangle_{U}|m_{-}\rangle_{L}+|m_{-}\rangle_{U}|m_{+}\rangle_{L})$
and $|\Phi^{+}\rangle_{UL}=\frac{1}{\sqrt{2}}
(|m_{+}\rangle_{U}|m_{+}\rangle_{L}+|m_{-}\rangle_{U}|m_{-}\rangle_{L})$
are two Bell states of the two ions. To express the evolution
clearly, we will consider the mixed state as the probabilistic
mixture of pure two-ion entangled states, i.e. the state
$|\Psi^{+}\rangle_{UL}$ with probability $F$ and the state
$|\Phi^{+}\rangle_{UL}$ with probability $1-F$. In the
$|\Psi^{+}\rangle_{UL}$ case, if $D_{u}$ fires the two ions will be
left in
$\frac{1}{\sqrt{2}}(|m_{+}\rangle_{U}|m_{-}\rangle_{L}+|m_{-}\rangle_{U}|m_{+}\rangle_{L})$
state. If $D_{l}$ fires the two ions will collapse into
$\frac{1}{\sqrt{2}}(|m_{-}\rangle_{U}|m_{+}\rangle_{L}-|m_{+}\rangle_{U}|m_{-}\rangle_{L})$
state. On the contrary, the $|\Phi^{+}\rangle_{UL}$ case only leads
to fire at $D_{u}$ with the two ions in
$|m_{-}\rangle_{U}|m_{-}\rangle_{L}$ state.

If we detect a photon at $D_{u}$, the two ions will be left in a
mixed state whose fidelity (with respect to $|\Psi^{+}\rangle_{UL}$
) is lower than the initial one. So we consider this result as
garbage. If $D_{l}$ registers one photon, the two ions are left in a
pure maximally entangled state $|\Psi\rangle_{UL}=\frac{1}{\sqrt{2}}
(|m_{-}\rangle_{U}|m_{+}\rangle_{L}-|m_{+}\rangle_{U}|m_{-}\rangle_{L})$
with probability $P'=\frac{F}{4}$. So long as the initial fidelity
satisfies $F>\frac{1}{2}$, the successful probability of the mixed
states case will be larger than the pure product states case. This
point can be understood easily. The pure states case starts from a
product state, but the mixed states case from a partially entangled
state. Naturally, the probability of the later case is larger than
the former one.

Then if the \emph{enclosure cavity} used in the pure product state
case has been added in the MZI of the mixed state case, the
scattering rate and the generation efficiency all can be enhanced.
Through analysis, we get that the successful probability of
getting pure maximally entangled states after iteration of the
process is $\frac{F}{3}$, which is larger than that of the one
round case $\frac{F}{4}$ with an increased scattering rate.

Compared to the previous generation schemes, our scheme has the
following  advantages:(1) The relative phase problem has been
avoided successfully in our scheme by using MZI, and the relative
phase in our scheme is adjusted to zero and will not change in the
process. The common phase of the state takes no effect on the
entanglement of the generated entangled states. (2)The photon we
want to detect is the input circular polarized photon, which makes
it easier to be registered than the scattered ones, because the
input photon has a better directionality than the scattered one.
(3)In the previous schemes, the BSA is a necessity, but in our
scheme, only two \emph{ordinary} BS are needed. So the current
scheme is simpler than the previous ones. (4)The simultaneity of the
two scattered photons is a main difficulty of the preceding schemes.
But in our scheme, the simultaneity will be satisfied naturally
because of the MZI.

\section{DISCUSSION}
\begin{figure}
\includegraphics[width=\columnwidth]{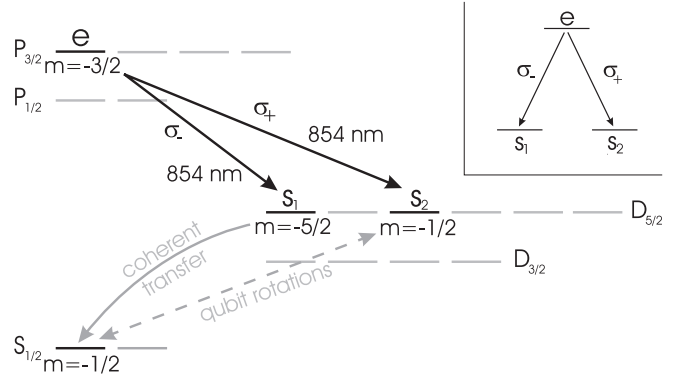}
\caption{Relevant levels of $^{40}$Ca$^{+}$ ions~\cite{simon}.}
\label{newlevel}
\end{figure}

After discussion on the generation scheme itself, we will consider
the feasibility of the current scheme. Singly positively charged
alkaline ions, which have only one electron outside a closed shell,
are commonly used in the quantum information experiments using
trapped ions~\cite{ion1,ion2}. Here we discuss a possible
implementation of our generation scheme using $^{40}$Ca$^{+}$ as
example. The relevant levels of $^{40}$Ca$^{+}$ has been depicted in
Fig. \ref{newlevel}~\cite{simon}.

$D_{5/2}$ and $D_{3/2}$ are two metastable levels of $^{40}$Ca$^{+}$
with lifetimes of the order of $1s$. $s_{1}$ and $s_{2}$ are two
sublevels of $D_{5/2}$ with $m=-5/2$ and $m=-1/2$, and this two
sublevels are coupled to $|e\rangle$ by $\sigma_-$ and $\sigma_+$
light at $854nm$. Here $e, S_{1}, S_{2}, S_{1/2}$ correspond to $e,
m_{-}, m_{+}, g$ in Fig. \ref{level}, respectively, i.e. we use the
$S_{1/2}$ as stable ground state, $S_{1}, S_{2}$ as two degenerate
metastable state and $P_{3/2}$ as excited state. Arbitrary
superposition state of this two degenerate metastable states can be
prepared by applying a laser pulse of appropriate length, and this
process can be realized in a few microsecond~\cite{ion4}. The
$^{40}$Ca$^{+}$ in state $S_{1}$ or $S_{2}$ can be excited into the
excited state $P_{3/2}$ by applying one $\sigma_-$ or $\sigma_+$
light at $854nm$. Then decay from $|e\rangle$ to $S_{1}, S_{2}$, to
$D_{3/2}$ and to $S_{1/2}$ are all possible. But the branching ratio
for $P_{3/2}\rightarrow D_{5/2}(854nm)$ versus $P_{3/2}\rightarrow
S_{1/2}(393nm)$ can be estimated as 1:30, giving $0.5 \times 10^7/$s
for the transition probability~\cite{ion2, simon}. So in most case,
the $^{40}$Ca$^{+}$ in the excited state will decay into the stable
ground state $S_{1/2}$. The detection of the internal states of
$^{40}$Ca$^{+}$ can be realized by using a cycling transition
between $S_{1/2}$ and $P_{1/2}(397nm)$~\cite{ion1,ion2}.

In section II and section III, we have discussed the effects of
the cavities enclosing the MZI (\emph{enclosure cavities}) on the
generation scheme, and next we will discuss the effects of the
resonant cavities surrounding the ions on the generation scheme.
To achieve directional emission of the photons from the ions, we
introduced an optical resonant cavity to surround each ion. We
will introduce a cavity on the $S_{1/2}$ to $P_{3/2}$ transition
to enhance the emission of the photons from atom transition
$P_{3/2}$ to $S_{1/2}$. Then the following two items will affect
the emission efficiency of the photon from the ions: (1) The
coupling between cavity mode and the $P_{3/2}\rightarrow
S_{1/2}(393nm)$ transition; (2) Decay from $P_{3/2}$ to $D_{5/2}$;
(3) Cavity decay. The probability $p_{cav}$ for a photon to be
emitted into the cavity mode after excitation to $|e\rangle$ can
be expressed as $p_{cav}=\frac{4 \gamma
\Omega^2}{(\gamma+\Gamma)(\gamma \Gamma + 4 \Omega^2)}$, where
$\gamma=4\pi c/F_{cav}L$ is the decay rate of the cavity,
$F_{cav}$ its finesse, $L$ its length,
$\Omega=\frac{D}{\hbar}\sqrt{\frac{hc}{2 \epsilon_0 \lambda V}}$
is the coupling constant between the transition and the cavity
mode, $D$ the dipole element, $\lambda$ the wavelength of the
transition, $V$ the mode volume (which can be made as small as
$L^2\lambda/4$ for a confocal cavity with waist
$\sqrt{L\lambda/\pi}$), and $\Gamma$ is the non-cavity related
loss rate~\cite{simon, decay}. From the discussion of
Ref.~\cite{simon}, the photon package is about 100ns, and such a
long coherence time makes it easy to achieve good overlap for the
wave function of the photon on the beam splitter.

When calculating the total efficiency of the generation scheme, we
must consider the following items:
\begin{itemize}
    \item The emission efficiency of photon: $p_{cav}$, which has included the cavity
    decay; To maximize the $p_{cav}$, we have chosen
    $F_{cav}=19000$, $L=3mm$. Then
    $\gamma=9.9\times10^{6}/s$, $p_{cav}=0.01$~\cite{simon};
    \item The efficiency of the photon detectors is expressed as
    $\eta$. Here we let the detection efficiency $\eta=0.7$, which is a level that can be reached within the current
    technology.
    \item Coupling the photon out of the cavities will introduce another error $\xi$, which can be modulated to be close to unit.

\end{itemize}

In addition, because the two ions have been placed on the MZI
symmetrically, the different transition times for the ions and the
consequent pulse broadening will affect the efficiency of the
scheme slightly. To complete the generation scheme, we suppose
that the state maker has held two ionic ensembles. After
considering the above factors, the total success probability can
be expressed as follow( considering the modified schemes as
example):

\begin{itemize}
    \item
    $P={\frac{F}{3}}\times{p_{cav}}\times{\eta}$
    for mixed state, that is to say, if we input photon with the
    rate of $5000/s$, we can get eight pairs of pure maximally entangled $^{40}$Ca$^{+}$
    ions per second for $F=0.7$.
    \item $P={\frac{2a^{2}(1-a^{2})}{3}}\times{p_{cav}}\times{\eta}$
    for product initial states, that is to say, if we input photon with the
    rate of $5000/s$, we can get five pairs of pure maximally entangled $^{40}$Ca$^{+}$
    ions per second for $a^{2}=0.7$.
\end{itemize}

From the experimental point of view, because the efficiency of the
current scheme would be greatly enhanced if there were enough
photons in the resonant system to induce stimulated emission from
the ions, we will input enough photons into the MZI
simultaneously. That is to say, the current scheme becomes more
realizable.

\section{CONCLUSION}

In conclusion, we have proposed an entanglement generation scheme,
which can entangle two ions by using MZI plus an optical
\emph{enclosure cavity}. Pure maximally entangled states can be
generated from either product states or mixed states. Single
photon detection can give us the signal indicating whether the
generation process succeeds or not. The added optical
\emph{enclosure cavity} enhances the generation efficiency.
Because the simultaneity problem inherent in the preceding schemes
does not appear in our scheme, ours is more realizable than the
previous ones.

\begin{acknowledgments}
This work is supported by Anhui Provincial Natural Science
Foundation under Grant No: 03042401, the Key Program of the
Education Department of Anhui Province under Grant No:2004kj005zd
and the Talent Foundation of Anhui University.
\end{acknowledgments}


\begin{thebibliography}{99}
\bibitem{en} A. Einstein, B. Podolsky, and N. Rosen, Phys. Rev. 47, 777 (1935).
\bibitem{hidden} D. M. Greenberger, M. Horne, A. Shimony, and A. Zeilinger, Am. J. Phys. 58, 1131 (1990).
\bibitem{cryp} A. K. Ekert, Phys. Rev. Lett. 67, 661 (1991).
\bibitem{comput} C. H. Bennett, and D. P. DiVincenzo, Nature 404, 247 (2000).
D. Gottesman, and I. L. Chuang, Nature 402, 390 (1999). J. Jones,
Nature 421, 28 (2003).
\bibitem{tele} C. H. Bennett, G. Brassard, C. Cr\'{e}peau, R. Jozsa, A. Peres, and W. K. Wootters, Phys. Rev. Lett. 70, 1895 (1993).
\bibitem{generation}E. Hagley, X. Maitre, G. Nogues, C. Wunderlich, M. Brune, J. M. Raimond, and S. Haroche, Phys. Rev. Lett. 79, 1 1997. A. Rauschenbeutel, G.
Nogues, S. Osnaghi, P. Bertet, M. Brune, J. M. Raimond, and S.
Haroche, Science 288, 2024 (2000).
\bibitem{ion} Q. A. Turchette, C. S. Wood, B. E. King, C. J. Myatt, D. Leibfried, W. M. Itano, C.
Monroe, and D. J. Wineland, Phys. Rev. Lett. 81, 3631 (1998).
\bibitem{nmr} N. A Gershenfeld and I. L Chuang, Science, 275, 350 (1997).
J. A Jones, M. Mosca, and R. H. Hansen, Nature 393, 344 (1998).
\bibitem{spdc} P. G. Kwiat, K. Mattle, H. Weinfurter, A. Zeilinger, A. V. Sergienko, and Y. Shih, Phys. Rev. Lett. 75, 4337 (1995).
A. G. White, D. F. V. James, P. H. Eberhard, and P. G. Kwiat,
Phys. Rev. Lett. 83, 3103 (1999). J.-W. Pan, D. Bouwmeester, M.
Daniell, H. Weinfurter, and A. Zeilinger, Nature (London) 403, 515
(2000).
\bibitem{zheng} S.-B. Zheng and G.-C. Guo, Phys. Rev. Lett. 85, 2392 (2000).
\bibitem{haroche} S. Osnaghi, P. Bertet, A. Auffeves, P. Maioli, M. Brune, J. M. Raimond, and S.
Haroche, Phys. Rev. Lett. 87, 037902 (2001).
\bibitem{bose} S. Bose, P. L. Knight, M. B. Plenio, and V. Vedral, Phys. Rev. Lett. 83, 5158 (1999).
\bibitem{browne} D. E. Browne, M. B. Plenio, and S. F. Huelga, Phys. Rev. Lett. 91, 067901 (2003).
\bibitem{cabrillo} C. Cabrillo, J. I. Cirac, P. Garc\'{\i}a-Fern\'{a}ndez, and P. Zoller, Phys. Rev. A 59, 1025 (1999).
\bibitem{dlm} L.-M. Duan, M. D. Lukin, J. I. Cirac \& P. Zoller, Nature, 414, 413 (2001).
\bibitem{dlm1} L.-M. Duan and H. J. Kimble, Phys. Rev. Lett. 90, 253601 (2003).
\bibitem{feng} X.-L. Feng, Z.-M. Zhang, X.-D. Li, S.-Q. Gong, and Z.-Z. Xu, Phys. Rev. Lett. 90, 217902 (2003).
\bibitem{lloyd} S. Lloyd, M. S. Shahriar, J. H. Shapiro and P. R. Hemmer, Phys. Rev. Lett. 87, 167903 (2001).
\bibitem{plenio} M. B. Plenio, S. F. Huelga, A. Beige, and P. L. Knight, Phys. Rev. A 59, 2468 (1999).
\bibitem{protsenko} I. E. Protsenko, G. Reymond, N. Schlosser, and P. Grangier, Phys. Rev. A 66, 062306 (2002).
\bibitem{simon} C. Simon and W. T. M. Irvine, Phys. Rev. Lett. 91, 110405 (2003).
\bibitem{nqi} X. Zhou, Z.-W. Zhou, M. J. Feldman, and G.-C. Guo, Phys. Rev. A 64, 044103 (2001).
X. Zhou, Z.-W. Zhou, G.-C. Guo, and M. J. Feldman, Phys. Rev. A,
64, 020101(R) (2001).
\bibitem{mepra} M. Yang, W. Song, and Z.-L. Cao, Phys. Rev. A 71,
012308 (2005).
\bibitem{purification} C. H. Bennett, G. Brassard, S. Popescu, B. Schumacher, J. A. Smolin, and W. K.
Wootters, Phys. Rev. Lett. 76, 722 (1996).
\bibitem{trapping} P. W. H. Pinkse, T. Fischer, T. P. Maunz, \& G. Rempe, Nature 404, 365 (2000).
J. Ye, D. W. Vernooy, and H. J. Kimble, Phys. Rev. Lett. 83, 4987
(1999).
\bibitem{pjw} J.-W. Pan, C. Simon, \v{C} Brukner \& A. Zeilinger, Nature 410, 1067 (2001).
\bibitem{pjw1} J.-W. Pan, S. Gasparoni, R. Ursin, G. Weihs \& A.Zeilinger, Nature 423, 417 (2003).
\bibitem{distillation} Z.-L. Cao, and M. Yang, J. Phys. B 36, 4245 (2003).
\bibitem{distillation1} Z.-L. Cao, M. Yang and G.-C. Guo Phys Lett A 308, 349 (2003).
\bibitem{ion1} F. Schmidt-Kaler, H. H\"{a}ffner, M. Riebe, S. Gulde, G. P. T. Lancaster, T. Deuschle, C. Becher, C. F. Roos, J. Eschner, and R. Blatt, Nature (London) 422, 408 (2003).
\bibitem{ion2} Ch. Roos, Th. Zeiger, H. Rohde, H. C. N\"{a}gerl, J. Eschner, D. Leibfried, F. Schmidt-Kaler, and R. Blatt, Phys. Rev. Lett. 83, 4713 (1999).
\bibitem{ion4} F. Schmidt-Kaler, S. Gulde, M. Riebe, T. Deuschle, A. Kreuter, G. Lancaster, C. Becher, J. Eschner, H. H\"{a}ffner and R. Blatt, J. Phys. B 36, 623 (2003).
\bibitem{decay} S. Haroche and J. M. Raimond, Adv. At. Mol. Phys. 20,350 (1985).
\end{thebibliography}
\end{document}